# Single exciton trapping in an electrostatically defined 2D semiconductor quantum dot

**Authors:**

Daniel N. Shanks[1], Fateme Mahdikhanysarvejahany[1], Michael R. Koehler[2], David G. Mandrus[3-5], Takashi Taniguchi[6], Kenji Watanabe[7], Brian J. LeRoy[1], and John R. Schaibley[1]*

**Affiliations:**

[1]Department of Physics, University of Arizona, 1118 E 4th St., Tucson, Arizona 85721, USA

[2]IAMM Diffraction Facility, Institute for Advanced Materials and Manufacturing, University of Tennessee, Knoxville, 2641 Osprey Vista Way, Knoxville, Tennessee 37920

[3]Department of Materials Science and Engineering, University of Tennessee, Knoxville, 1508 Middle Dr., Knoxville, Tennessee 37996, USA

[4]Materials Science and Technology Division, Oak Ridge National Laboratory, 1 Bethel Valley Rd., Oak Ridge, Tennessee 37831, USA

[5]Department of Physics and Astronomy, University of Tennessee, Knoxville, 1408 Circle Dr., Knoxville, Tennessee 37996, USA

[6]International Center for Materials Nanoarchitectonics, National Institute for Materials Science, 1-1 Namiki, Tsukuba 305-0044, Japan

[7]Research Center for Functional Materials, National Institute for Materials Science, 1-1 Namiki, Tsukuba 305-0044, Japan

**\*Corresponding Author:** John R. Schaibley, johnschaibley@email.arizona.edu

**Abstract:**

Interlayer excitons (IXs) in 2D semiconductors have long lifetimes and spin-valley coupled physics, with a long-standing goal of single exciton trapping for valleytronic applications. In this work, we use a nano-patterned graphene gate to create an electrostatic IX trap. We measure a unique power-dependent blue-shift of IX energy, where narrow linewidth emission exhibits discrete energy jumps. We attribute these jumps to quantized increases of the number occupancy of IXs within the trap and compare to a theoretical model to assign the lowest energy emission line to single IX recombination.

**Main text:**

WSe$_2$-MoSe$_2$ heterostructures are known to host spatially indirect excitons, with long (>1ns) lifetimes due to the spatial separation of the electron and hole in different transition metal dichalcogenide (TMD) layers [1–7]. Signatures of multiple IXs occupying a single trap have been observed in naturally occurring IX traps [8], and nano-pillar based strain induced traps [9]. In previous work, we demonstrated nanoscale trapping of IXs using nano-patterned graphene to induce a spatially varying electric field, which interacts with the permanent out-of-plane dipole moment of IXs to create an electrostatically defined quantum dot (QD) for IXs but were unable to resolve single IX trapping [10]. In this work, we report discrete energy emission lines with a unique power dependence coming from an electrostatic IX QD, a signature of few-IX interactions and single-IX trapping. These traps are advantageous over other trapping methods involving strain or moiré traps, due to their deterministic placement in a lithographically defined process and 100 meV energy tunability by applied gate voltage, with potential for applications as optically-active QD systems to serve as single photon emitters or quantum spin-photon interfaces [11,12].

Figures 1(a)-1(b) show the layout of the electrostatic IX QD device, with an optical image of the device shown in the Supplemental Material [13] (Fig. S1). We used mechanical exfoliation and dry transfer to fabricate an hBN encapsulated WSe$_2$-hBN-MoSe$_2$ heterostructure, with top and bottom few-layer-graphene (FLG) gates. The thin hBN separator layer between the TMDs strongly reduces the induced moiré potential energy landscape and increases the dipole moment of IXs, allowing for larger energy tunability of IXs [14–16]. TMD layers were angle aligned, with their crystal axes determined by second harmonic generation measurements [17–19]. The TMDs in this device have a near zero degree twist angle (R-type), determined by the energy and circular polarization of PL measurements on an adjacent region of the sample without the hBN spacer [16]. The top layer of hBN is chosen to be thin to allow for close proximity to the patterned top graphene layer. After the transfer step, we performed high resolution (100 kV) electron beam lithography and O$_2$ reactive ion etch to pattern the graphene top gate, producing an oblong oval hole, measured to be 30 x 70 nm by atomic force microscopy (AFM), shown in Fig. 1(c). Previous studies have shown that hBN encapsulation of TMD layers preserves their optical properties under e-beam lithography exposure and O$_2$ etching, and that hBN is resistant to O$_2$ etching, which prevents the creation of processing-produced defects or trapping [20–22]. A second round of electron beam lithography and thermal evaporation of Cr/Au was performed to electrically contact to the sample.

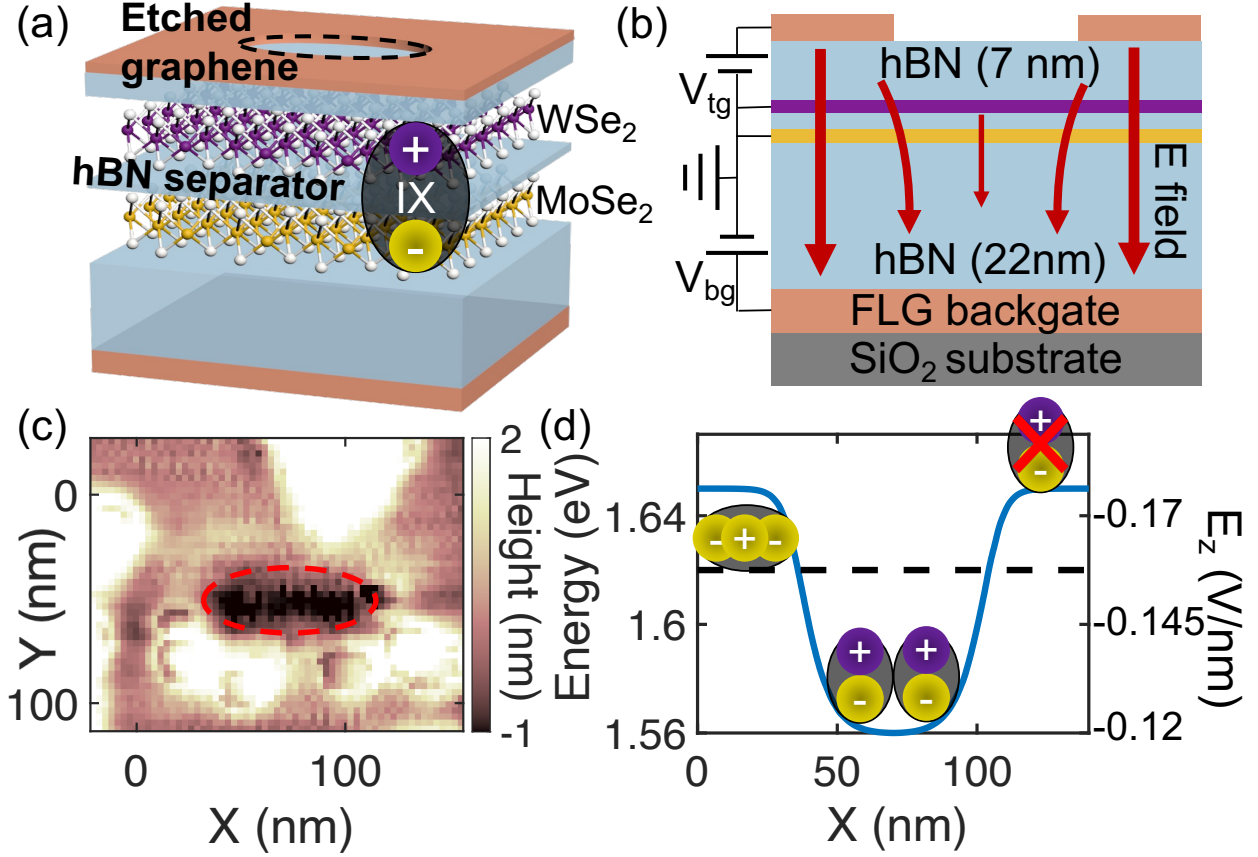

FIG. 1. Layout of the IX trapping structure. (a) Depiction of the MoSe$_2$-WSe$_2$ heterostructure, with a thin hBN spacer layer, encapsulated in hBN with a patterned graphene top gate and FLG back gate. (b) Depiction of the spatially varying electric field through the heterostructure. (c) AFM topography of the etched graphene hole, with the etched area outlined by the dashed red line. (d) Diagram of the inter- and intra-layer exciton energy when the electric field is set such that the field away from the hole is -0.18 V/nm. The IX energy shifts with electric field due to the permanent out-of-plane dipole moment, while the intralayer exciton energy is unchanged.

Figure 1(d) shows the available energy states for inter- and intra-layer excitons in a strong, spatially varying applied electric field underneath the etched graphene. The electric field profile created by the nano-patterned graphene was simulated in a 3D COMSOL model, matching the shape of the graphene etch in the model to the AFM topography. The COMSOL simulations used an electrostatic model using known values for the in-plane and out-of-plane dielectric constants of the hBN, and hBN top layer thickness that matched the AFM data of the individual 2D layers [23]. Voltage was applied at the top graphene-hBN interface, while the top surface of the TMD heterostructure was grounded. The minimum and maximum IX energies and field strengths are set to match the experimental data in Fig. 2.

With no applied electric field, the IX energy both underneath and away from the etched graphene is measured to be 1.4 eV, consistent with measurements on other hBN-separated MoSe$_2$-WSe$_2$ heterostructures [14–16]. When the electric field is applied in the opposite direction of the IX dipole moment, the IX energy increases proportionally to the strength of the local electric field by H = −**p** · **E**, where H is the IX energy, **p** is the permanent dipole moment of the IX, and **E** is the spatially varying electric field coming from the etched graphene. We calculate the electric field applied by the gate voltages using the three-plate capacitor model for the hBN-separated TMD layers, described in *Unuchek et al.* [14]. For a strong enough applied electric field, the IX energy can be tuned above the energy of the lowest available intralayer ($MoSe_2$) charged exciton state. However, because the electric field underneath the etched graphene is weaker, for a certain range of gate voltages, the lowest available energy state in this small area remains the trapped IX state.

Photoluminescence (PL) measurements were performed at 6 K in an optical cryostat (Montana Instruments), using a 670 nm diode laser and a grating-based spectrometer with a cooled CCD camera. We used a 0.6 NA 40x objective and a confocal pinhole which resulted in a collection diameter of 1.5 μm on the sample, aligned to the nanopatterned graphene oblong hole. Figure 2(a) shows gate dependent photoluminescence (PL), where we applied a topgate/backgate voltage ratio equal to the relative thickness of the top and bottom hBN, $\alpha$ = -$V_{bg}$/$V_{tg}$ = 3.2, to apply an electric field while keeping the heterostructure overall charge neutral [24]. As the strength of the electric field increases, the PL energy from free IXs in the area surrounding the removed graphene and trapped IXs underneath the removed graphene increases linearly as expected.

At low electric field (less than -0.15 V/nm), the free IX PL signal shifts at a rate of 1.4 eV/(V/nm), leading to a measured dipole moment of 1.4 nm. At an electric field near -0.155 V/nm, the free IX energy is tuned to 1.62 eV, equal to the $MoSe_2$ intralayer trion energy. We note that the prevalence of the $MoSe_2$ trion at 1.62 eV, and relatively weak signal from the $MoSe_2$ neutral exciton at 1.65 eV is consistent with the overall bilayer heterostructure being charge neutral, while the individual $MoSe_2$ and $WSe_2$ layers are doped equally and oppositely. At electric fields more negative than -0.16 V/nm, the IX energy would be higher than the $MoSe_2$ intralayer trion energy, but there is no observed PL signal from free IXs in this electric field regime. Additionally, the brightness from intralayer trions increases above this field, indicating that the population of holes in the $MoSe_2$ layer that would undergo charge transfer into the $WSe_2$ to create IXs no longer do so, resulting in a higher population of $MoSe_2$ intralayer particles. The trapped IX PL signal appears between 1.5 eV to 1.62 eV, and shows multiple, discrete narrow lines, with linewidths down to 0.9 meV. These lines can be attributed to electrostatically trapped IXs, as the reduced dipole shift is a signature of the weaker electric field experienced by these excitons underneath the removed graphene [10].

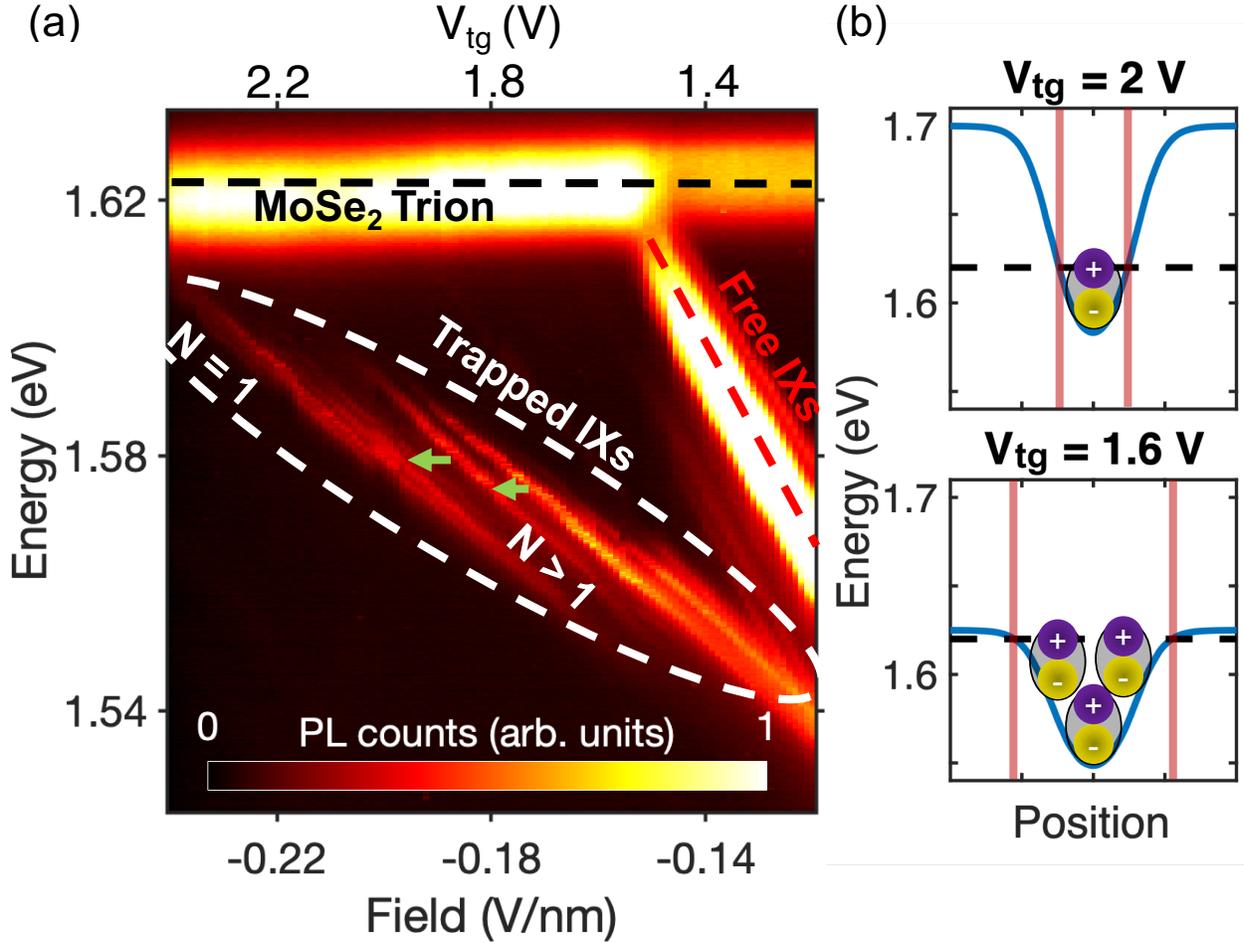

FIG. 2. (a) PL from a single IX QD as a function of electric field, with excitation power 125 nW. The trapped IX PL shows discrete narrow emission lines, and energy shifts with electric field shown by green arrows. The electric field axis refers to electric field strength away from the etched graphene. (b) Exciton energy near the etched graphene for varying applied voltage. Blue line indicates IX energy, black dashed line indicates MoSe$_2$ trion energy. Red lines mark the intersection, showing that with increasing applied voltage, there is less area where excitons will fall to the trapped IX state.

With increasing excitation power, the trapped IX signal is observed at higher energy (see Fig. S2). The higher laser power creates a higher density of excitons in the vicinity of the trap, resulting in higher trapped IX population. The observed PL energy shift can be attributed to IX-IX repulsion between IXs in the same trap, which comes from two different sources. One source is Coulomb dipolar repulsion energy $E_{p-p}(\Delta r)$ due the repulsion of two dipoles at a distance $\Delta r$ with their orientation locked to be parallel, previously observed in both semiconductor quantum well systems and TMD heterostructures [8,9,25–27]. The other source is IXs pushing themselves away from one another, up the side-walls of the trap to an area of higher electrical potential energy $V_{IX}(x,y)$ which depends on each IX position coordinates within the trap $(x,y)$.

As the strength of the electric field increases, the trapped IX signal exhibits discrete shifts to lower energy states, indicating lower population of the trap. As the magnitude of the applied voltage is increased, there is less area on the sample where the IX state is lower energy than the MoSe$_2$ intralayer trion energy. Thus, there is less area where excitons fall to the IX state, producing a lower average occupation of the IX trap, shown by Fig. 2(b). At the strongest applied field (<-0.22 V/nm), when the lowest spatial IX energy in the center of the trap meets the MoSe$_2$ trion energy, the lowest PL emission line represents emission from a single trapped IX, as the system crosses the limiting case towards zero-population of trapped IXs. We observe qualitatively similar discrete energy shifts in two other electrostatic QD structures (see Fig. S3, and references [28–30] therein).

We note a 10% increase in the slope of the dipole energy shift of the high population states (-0.16 V/nm < E$_z$ < -0.12 V/nm) compared to the low population states (E$_z$ < -0.18 V/nm) in Fig 2a. This change in dipole energy shift comes from IXs in high population states being pushed up the side walls of the trap, where the fractional electric field strength is larger. This confirms the electrostatic nature of the IX trap and rules out strain as the primary method of trapping, which would not produce a dipole shift which changes with population of the trap (see Fig. S4-S5).

In order to characterize the discrete energy transitions, excitation power dependent PL was performed for low powers on the trapped IX signal, shown in Fig. 3(a). At the lowest measured excitation power, 8 nW, the lowest energy peak is most prominent and centered at $1561.44 \pm 0.03$ meV, with a small signal from a secondary peak at $1563.16 \pm 0.05$ meV. As the power is increased, the number of IXs in the QD increases. At 18 nW excitation power, the lowest energy PL peak becomes dimmer, and the second peak becomes the strongest, with a rise of a third peak at $1564.2 \pm 0.1$ meV. We attribute these three transitions to the lowest possible numbers of IX occupancy in the trap, N = 1→0, 2→1, and 3→2 transitions of IX population. Discrete power-dependent energy transitions such as these have been previously observed in MoSe$_2$-WSe$_2$ heterostructures [8,9], III-V quantum dots [31,32], and electrostatically trapped excitons in III-V coupled quantum wells [33] and are attributed to quantized numbers of IXs within these nanoscale traps. While single photon emission is expected from a singly trapped exciton, a photon antibunching experiment was not performed here due to weak emission intensity from the sample.

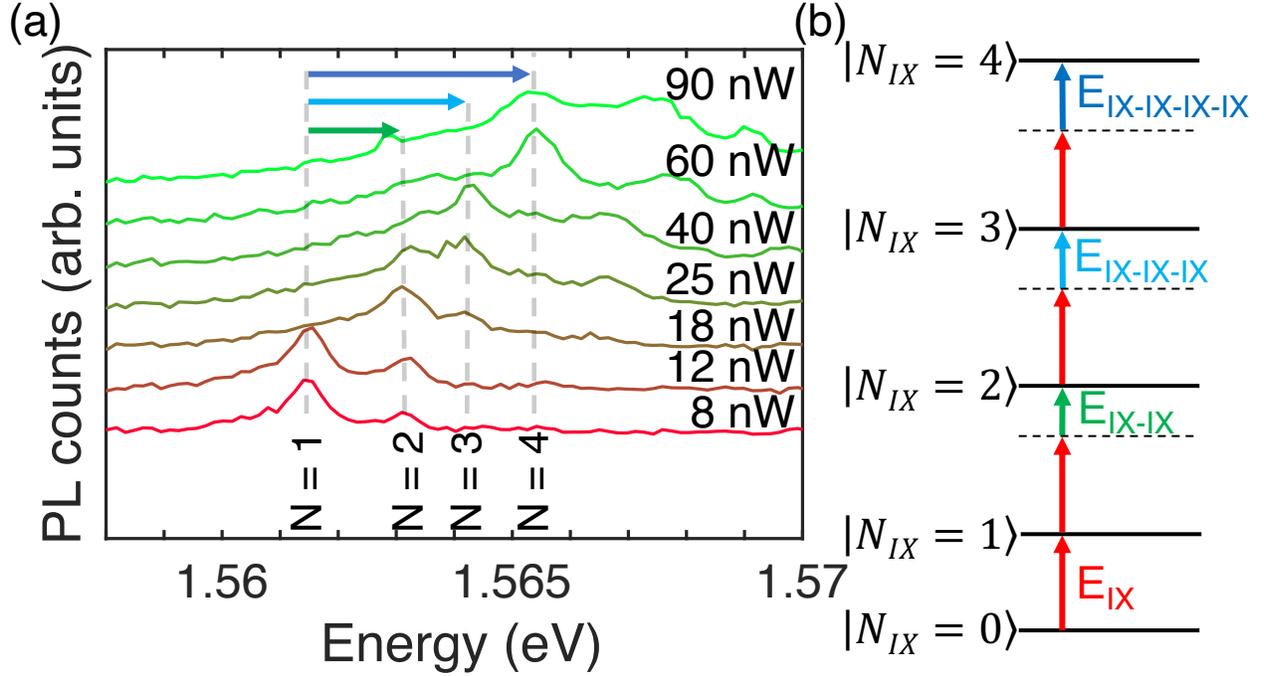

FIG. 3. Power dependent PL on an IX QD. (a) PL for various laser excitation powers between 8 and 90 nW, with the gate voltage $V_{tg} = -1/\alpha * V_{bg} = 1.8$ V, corresponding to an electric field strength away from the hole of -0.18 V/nm. Grey dashed lines indicated discrete emission peaks, green and blue arrows represent measured IX-IX repulsion energies. (b) Energy diagram of multi-exciton states confined within the same QD. $E_{IX}$ represents the lowest available energy state for a single IX at the applied electric field in the QD. $E_{IX}+E_{IX-IX}$ represents the additional energy to add a second IX to the same QD. $E_{IX}+E_{IX-IX-IX}$ ($E_{IX}+E_{IX-IX-IX-IX}$) represents the additional energy to add a third (fourth) IX to the QD.

Figure 3(b) shows the energy diagram and measured IX-IX repulsion energies for these transitions. We measure these experimental values by fitting the PL data to a multiple peak fit and taking the average of the center of these fits, with the uncertainty as the standard deviation of the center wavelength between different excitation power scans, with some representative fits shown in Fig. S6. Increasing the excitation power further causes the third peak to become the most prominent, with the eventual prominence of a fourth peak, until the linewidth broadens such that individual lines are no longer resolvable, representing a continuum of exciton states for excitation power above 5 μW, shown in Fig. S7.

We consider other possible origins of the discrete jumps in the trapped IX PL signal, including charging of the QD with single electrons, and quantum confinement giving rise to quantized energy states. We rule out charging of the QD by the unique power dependent blue shift of the IX peak in Fig. 3(a). In order to rule out the possibility of quantum confinement effects we estimated the quantum energy level spacing (see Supplementary note I, and reference [34] therein), and found a lowest energy transition of 0.45 meV, four times lower than the lowest experimentally observed transition. Additionally, the quantum confinement would not produce the unique power dependence observed, where the lowest energy peak disappears with increasing power, and thus can be ruled as the source of these energy transitions.

In these low power PL measurements, the dipole-dipole repulsion energy between IXs is on the order of 1-4 meV, consistent with the energy scale of multi-IX signatures in other reports [8,9]. We note that these previous results only report multi-exciton states up to N = 3 [8] and N = 5 [9], whereas our electrostatic trap produces a significantly higher upper limit of multi-exciton states, eventually producing a continuum-like state. In order to verify that the observed power-induced blue shift is indeed due to the lowest population IX transitions, we compare our data to a theoretical model for IXs. We calculated the expected IX repulsion energy by simulating the minimum energy positions of IXs within the spatially varying potential, balancing dipole-dipole repulsion that pushes IXs away from one another and the potential energy gradient induced by the etched graphene that pushes them together. This potential energy landscape for IXs is proportional to the electric field, which is determined by the COMSOL model of the structure. The dipole-dipole repulsion $E_{p-p}(\Delta r)$ is modeled using the Bilayer Keldysh Potential [35] as,

$$E_{p-p}(\Delta r) = 2 * \left( V_k(\Delta r) - V_k\left(\sqrt{\Delta r^2 + d^2}\right) \right) (1)$$

Where $\Delta r$ is the in-plane distance between IXs and d is the measured vertical electron-hole separation of the excitons, 1.4 nm. The first term in parentheses in equation 1 represents the repulsive electron-electron and hole-hole potential energy between charges in the same layer, and the second term represents the attractive potenial energy between electrons and holes in different IXs. $V_k(x)$ is the Bilayer Rytova-Keldysh potential [35,36] which has been shown to model the electric potential created by charges in TMDs taking into account dielectric screening of the environment [37] (see Supplementary note II, and reference [38] therein).

We minimize the total energy of the N-exciton system,

$$E_{N-IX} = \frac{1}{2} \sum_{i,j} E_{p-p}\left(\Delta r_{ij}\right) + \sum_i V_{IX}(x_i, y_i) \ (2)$$

Where $E_{p-p}$ is the dipolar repulsion term summed over all *(i,j)* exciton pairs with a 1/2 term to account for double-counting, and $V_{IX}$ is the discretized spatial IX energy dependence of each individual IX given by the COMSOL electric field model. The minimum $V_{IX}$ = 1.56 eV was set to match Fig. 3(a), and the maximum $V_{IX}$ = 1.65 eV away from the etched graphene was determined by extrapolating the energy of free IXs to their value at the electric field = -0.18 V/nm, given the dipole shift of the free IX energy in Fig. 2(a). We use a numerical solver to minimize the IX energy, which was stable and reproducible for multiple IX initial locations within the trap. The quantum exchange integral term, which is repulsive for IXs in the same valley and zero for IXs in opposite valleys, was neglected in this simulation, as we are comparing our model to the lowest experimentally observed energy splitting, and the exchange energy term will only increase this energy splitting. Figure 4 shows the locations of IXs in some of the multi-IX simulations, and Table 1 shows the calculated multi-IX energies from the simulation in Fig. 4, compared to the experimental values in Fig. 3. $E_{IX-IX}$ = $E_{2-IX}$-2($E_{IX}$), where $E_{IX}$ is the ground state energy of a single IX in the simulation, 1.56 eV, and $E_{2-IX}$ is the minimum energy for two excitons in the trap as given by equation 2. $E_{IX-IX-IX} = E_{3-IX} - E_{2-IX} - E_{IX}$ is the additional potential energy from the third exciton in the trap, and $E_{IX-IX-IX-IX} = E_{4-IX} - E_{3-IX} - E_{IX}$.

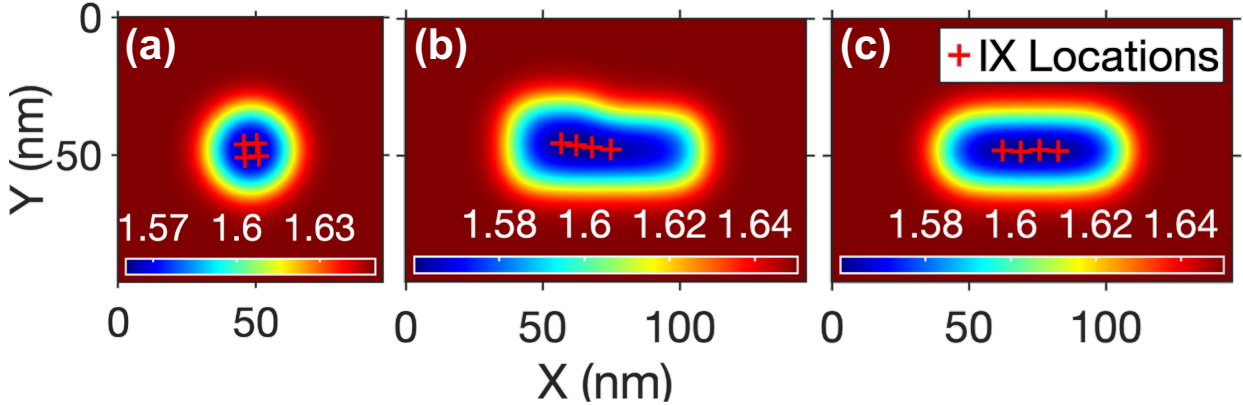

FIG. 4. Simulated locations for the 4-exciton state for a circular trap of diameter 35 nm (a), an oblong shaped hole whose dimensions most closely match that of the AFM topography (b), and an oval hole of 30 x 70 nm (c). Colorbar shows $V_{IX}(x,y)$, whose spatial profile is determined from the COMSOL simulation of the electric field underneath the etched graphene. The minimum and maximum $V_{IX}$ are set based on the measured data from Figs. 2 and 3.

|  | $E_{IX-IX}$ | $E_{IX-IX-IX}$ | $E_{IX-IX-IX-IX}$ |
|---|---|---|---|
| Experiment | $1.73 \pm 0.05$ | $2.8 \pm 0.1$ | $3.9 \pm 0.1$ |
| Circular etch | 3.9 | 5.6 | 7.9 |
| Oblong etch | 1.3 | 2.5 | 3.7 |
| Oval etch | 0.7 | 1.5 | 2.5 |

TABLE I: Experimental multi-IX energies from Fig. 3 compared to simulated multi-IX energies from Fig. 4. All units are meV.

The multi-IX energies from a circular graphene etch in Fig. 4(a) overestimate the measured multi-IX energy levels, while the oval hole in 4(c) slightly underestimates the observed multi-IX energies. We note that small changes in the etched graphene shape can produce significant changes in the IX repulsion energy. Upon careful investigation of the AFM topography, we observe that one side of the graphene etch is slightly wider than the other (see Fig. S8). Taking this into account, we modify our COMSOL simulation of the etched graphene shape to more closely represent the asymmetric shape determined by AFM topography, whose electric field profile is shown in Fig. 4(b). In this simulation, the multi-exciton energies are within 25% of the experimentally measured values. Alternatively, this potential energy landscape would also accurately represent the oval graphene etch in Fig. 4(c), with a small change in the potential energy across the length of the hole due to randomly induced potential energy variation across the sample described in *Mahdikhanysarvejahany et al.* [16], possibly due to strain [39].

In conclusion, we observe evidence of single and few IX trapping in an electrostatically defined 2D semiconductor quantum dot. Our approach using patterned graphene gates is highly customizable and results in location specific placement of the quantum dot as well as highly tunable emission energies via gate voltage. While there has been recent progress in alternative methods to lithographically define quantum dots in TMD systems that involve direct patterning of the semiconducting material [20,21,40], our approach leaves the TMD crystal intact, neutralizing the effect of edge states. Future studies will involve QD-trapped excitons in a magnetic field to break valley degeneracy, creating a valley-based two-level system, or investigating the ability to controllably charge the quantum dot with single electrons to realize a highly deterministic spin-photon interface [41], which can be potentially be interfaced with nanoplasmonic waveguides to realize single photon transistors [42] and quantum repeaters [11].

## Acknowledgements:

The authors acknowledge useful conversations with Dr. Rolf Binder.

JRS and BJL acknowledge support from the National Science Foundation Grant. Nos. ECCS-2054572 and DMR-2003583 and the Army Research Office under Grant no. W911NF-20-1-0215. JRS acknowledges support from Air Force Office Scientific Research Grant Nos. FA9550-18-1-0390 and FA9550-21-1-0219. BJL acknowledges support from the Army Research Office under Grant no. W911NF-18-1-0420 and the National Science Foundation Grant ECCS- 2122462. DGM acknowledges support from the Gordon and Betty Moore Foundation's EPiQS Initiative, Grant GBMF9069. K.W. and T.T. acknowledge support from JSPS KAKENHI (Grant Numbers 19H05790, 20H00354 and 21H05233). Plasma etching was performed using a Plasmatherm reactive ion etcher acquired through an NSF MRI grant, award no. ECCS-1725571.

## Author Contributions:

DNS, JRS, and BJL conceived the project. JRS and BJL supervised the project. DNS and FM fabricated the structures. DNS modelled the structures and performed the experiments. DNS analyzed the data with input from FM, JRS and BJL. MRK and DGM provided and characterized the bulk $MoSe_2$ and $WSe_2$ crystals. TT and KW provided hBN crystals. DNS, JRS and BJL wrote the paper. All authors discussed the results.

## Data availability:

All data needed to evaluate the conclusions in the paper are present in the paper or the Supplemental material. Source or additional data related to this paper may be requested from the corresponding authors.

## Competing Interests:

The authors declare no competing interests.

**Supplemental Material for**

Single exciton trapping in an electrostatically defined 2D semiconductor quantum dot

Daniel N. Shanks[1], Fateme Mahdikhanysarvejahany[1], Michael R. Koehler[2], David G. Mandrus[3-5], Takashi Taniguchi[6], Kenji Watanabe[7], Brian J. LeRoy[1], and John R. Schaibley[1]*

**Affiliations:**

[1]Department of Physics, University of Arizona, 1118 E 4th St., Tucson, Arizona 85721, USA

[2]IAMM Diffraction Facility, Institute for Advanced Materials and Manufacturing, University of Tennessee, Knoxville, 2641 Osprey Vista Way, Knoxville, Tennessee 37920

[3]Department of Materials Science and Engineering, University of Tennessee, Knoxville, 1508 Middle Dr., Knoxville, Tennessee 37996, USA

[4]Materials Science and Technology Division, Oak Ridge National Laboratory, 1 Bethel Valley Rd., Oak Ridge, Tennessee 37831, USA

[5]Department of Physics and Astronomy, University of Tennessee, Knoxville, 1408 Circle Dr., Knoxville, Tennessee 37996, USA

[6]International Center for Materials Nanoarchitectonics, National Institute for Materials Science, 1-1 Namiki, Tsukuba 305-0044, Japan

[7]Research Center for Functional Materials, National Institute for Materials Science, 1-1 Namiki, Tsukuba 305-0044, Japan

**Corresponding Author:** John R. Schaibley, johnschaibley@email.arizona.edu

Supplementary Note I:

To estimate the quantum confinement energy splitting within the electrostatic trap, we use a spatially discretized time-independent Schrödinger solver to find the eigenmodes of the electrostatically induced potential energy for single particle states. Starting from the time-independent Schrödinger equation:

$$E_j \psi_j(x) = \hat{H}_j \psi_j(x) = -\frac{\hbar^2}{2m}\frac{d^2\psi_j(x)}{dx^2} + V(x)\psi_j(x)$$

We introduce a discretized physical space with spatial index $n$ for both the wavefunction and potential energy, with distance between grid points $\Delta x$. We use the mass of the interlayer exciton m = 1.2 $m_0$ [34]. We replace the second derivate of the wavefunction with a spatially discretized version:

$$\frac{d^2\psi_n}{dx^2} \rightarrow \frac{\psi_{n+1} - 2\psi_n + \psi_{n-1}}{\Delta x^2}$$

leading to a discrete matrix form of the Hamiltonian:

$$H_{\mu,\nu} = -\frac{\hbar^2}{2m\Delta x^2}\left[\delta_{\mu,\nu+1} + \delta_{\mu,\nu-1}\right] + \left[\frac{\hbar^2}{m\Delta x^2} + V_\mu\right]\delta_{\mu,\nu}$$

Where $V_\mu$ is the discretized potential energy from the COMSOL model of the structure along the long direction of the hole (see Fig. 1d), which provides less spatial confinement and will determine the lowest energy splitting between quantum confined states. We numerically solve for the eigenvalues of the system

$$H_{\mu,\nu}\psi_\nu = E\,\psi_\nu$$

to obtain the energy levels.

Supplementary Note II:

The bilayer Keldysh potential energy $V_k(x)$, [35] for charges a distance $x$ apart from one another, is a modified version of the Rytova-Keldysh potential [36].

$$V_k(x) = \frac{e^2}{4\pi\epsilon_0\epsilon_r} * \frac{\pi}{2r_0}\left(\left(H_0\left(\frac{\epsilon_r x}{r_0}\right) - Y_0\left(\frac{\epsilon_r x}{r_0}\right)\right)\right) \quad (2)$$

Where $\epsilon_r = \sqrt{\epsilon_\parallel * \epsilon_\perp}$ is the effective dielectric environment seen by the carriers [4]. The characteristic screening length in the bilayer structure, $r_0$, is the sum of the characteristic screening lengths of monolayer WSe$_2$ and MoSe$_2$ [35], obtained from the 2D polarizabilities in *Berkelbach et al.* [38]. H$_0$ and Y$_0$ are the Bessel function of the second kind and Struve functions.

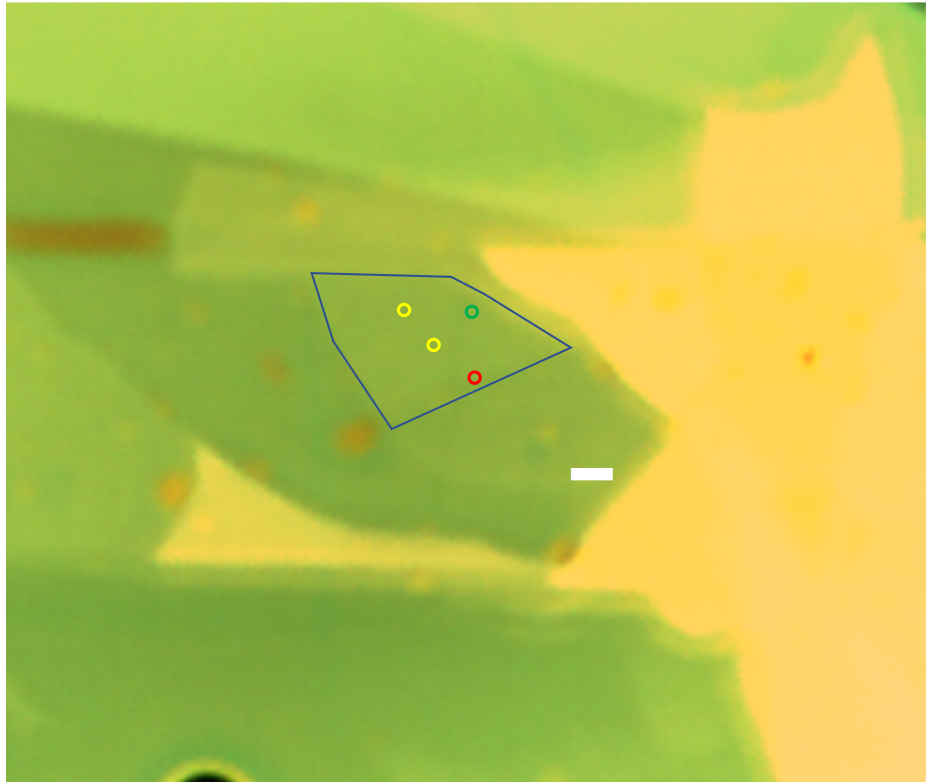

FIG. S1. Optical Image of the device. Blue lines show the outline of the hBN-separated TMD heterostructure. Green circle shows location of QD in main text. Yellow circles show alternate QD sites, with data shown in Fig S3(b) and (c). Red circle shows location of QD that showed very weak trapped IX signal, with no transitions between population states. White line shows scale bar, 1 μm.

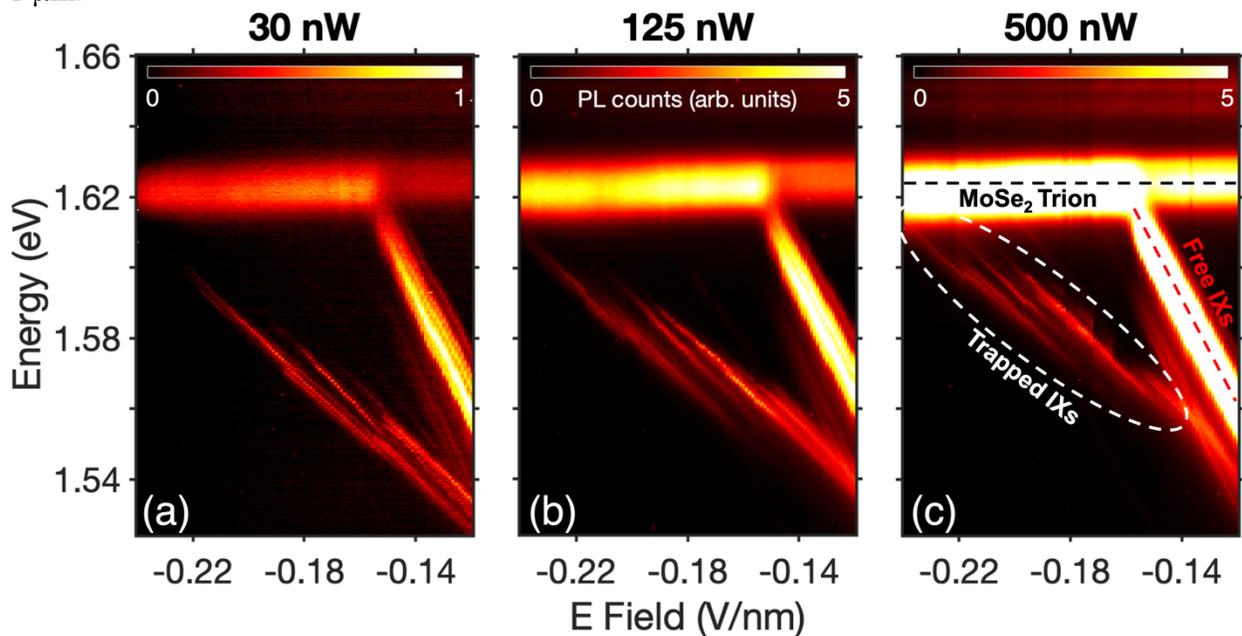

FIG. S2: Confocal PL gate maps for various excitation power, showing an increase in the measured trapped IX energy with higher excitation power.

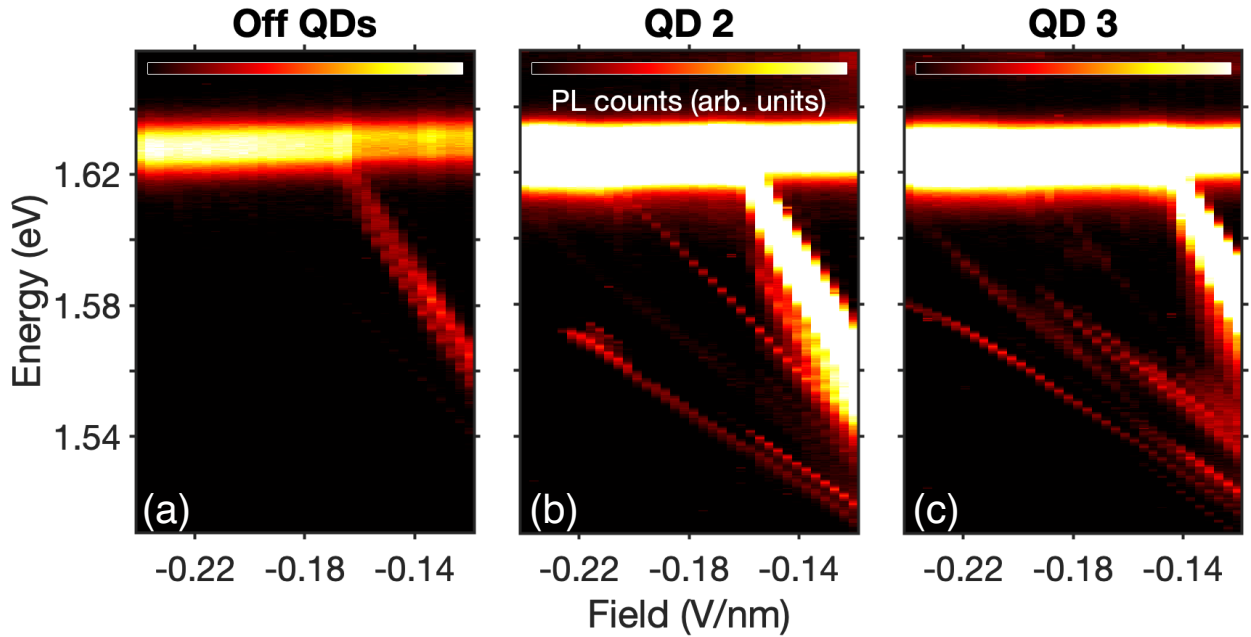

FIG. S3. (a) Confocal gate dependent PL on the hBN-separated TMD heterostructure off the lithographically defined QDs, showing the free IX signal, the MoSe$_2$ trion signal, and no signature of electrostatically trapped IXs. (b-c) PL from alternate QD sites with larger areas of removed graphene, 70 x 70 nm compared to the 30 x 70 nm structure studied in the main text, which provides weaker electrostatic trapping. These sites show narrow emission lines with reduced dipole shift compared to the main IX peak, the primary indicator of electrostatically trapped IXs. The trapped IX signal shows discrete energy transitions to lower energy with increasing applied voltage, as in the data from the main QD. These QDs show fewer energy shifts than the main QD in Figure 2, indicating lower average exciton population than the main QD structure for the same excitation power and gate voltage. We attribute this difference to the local strain environment surrounding the traps, which funnels IXs towards or away from different traps at different rates [28,29], possibly due to the presence of nano bubbles [30,39]. These alternate QDs also show some trapped IX signals with higher energy and dipole shift, indicating strain trapped excitons along the side walls of the electrostatic trap.

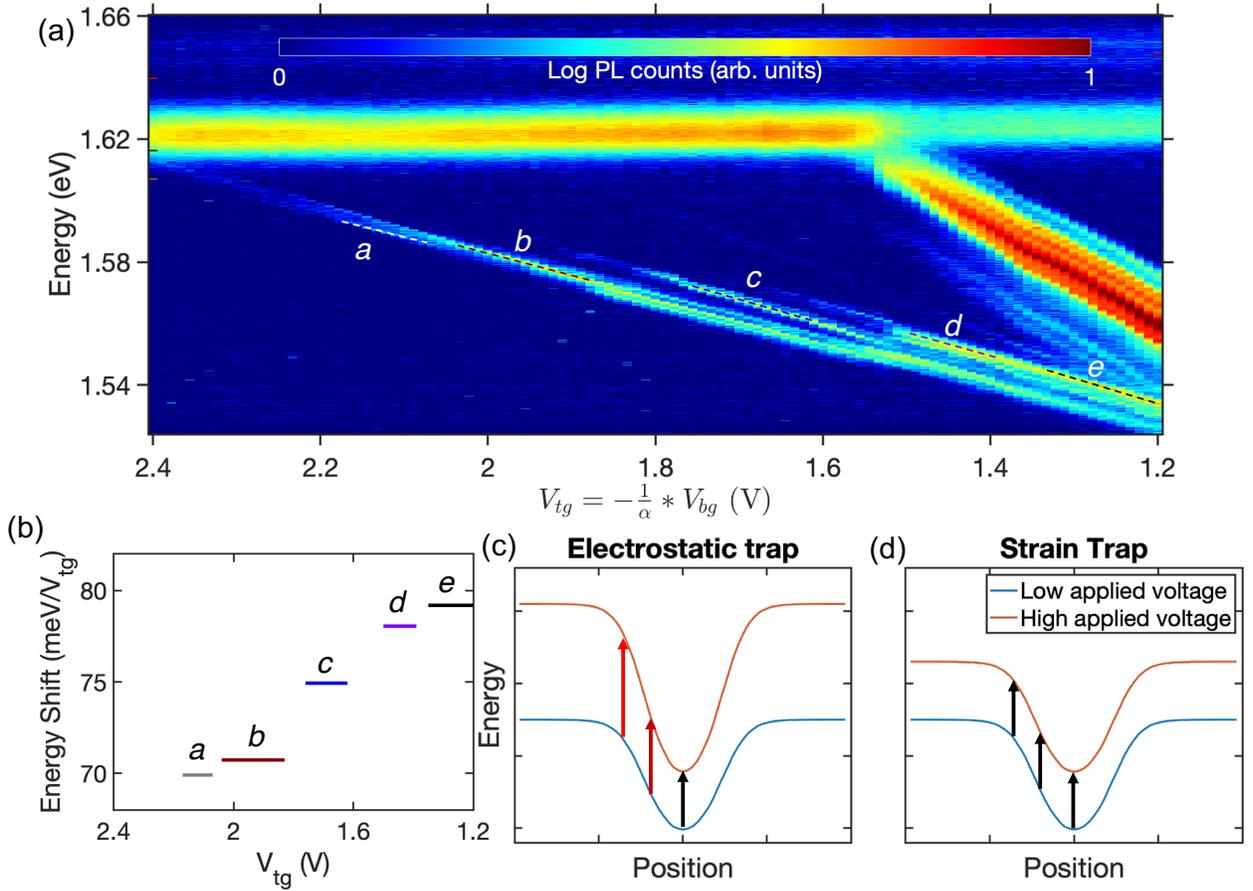

FIG. S4. Dipole shift of the trapped IX signal at low excitation power. (a) PL Data from Fig. S2(a) in units of applied voltage, with IX PL lines representing the highest population IX states for several voltage ranges marked as dashed lines. (b) Absolute value of IX energy shift per applied voltage for PL lines marked in (a), showing trend of increasing dipole shift with increased population of the trap. Dipole shifts are calculated by taking the change in the maximum emission energy of a chosen peak divided by the change in applied topgate voltage. (c) Depiction of spatial trapping potential in an electrostatic trap defined by nano-patterned graphene, where excitons pushed up the sidewalls of the trap in higher occupation states will experience a greater dipole-energy shift as a function of applied voltage due to the increased electric field strength along the side-walls of the trap. (d) Depiction of a fixed-depth strain trap in a spatially constant electric field, where excitons in higher population states show the same dipole shift as those in low population states, as observed in [8]. The change in dipole shift for different occupation states in (b) confirms that the induced electrostatic potential is the dominant source of trapping within the IX energy landscape, as a strain-induced IX trap [9], which would otherwise become the strongest source of trapping when the moiré potential is removed [16], would not produce the observed change of dipole shift with occupation level.

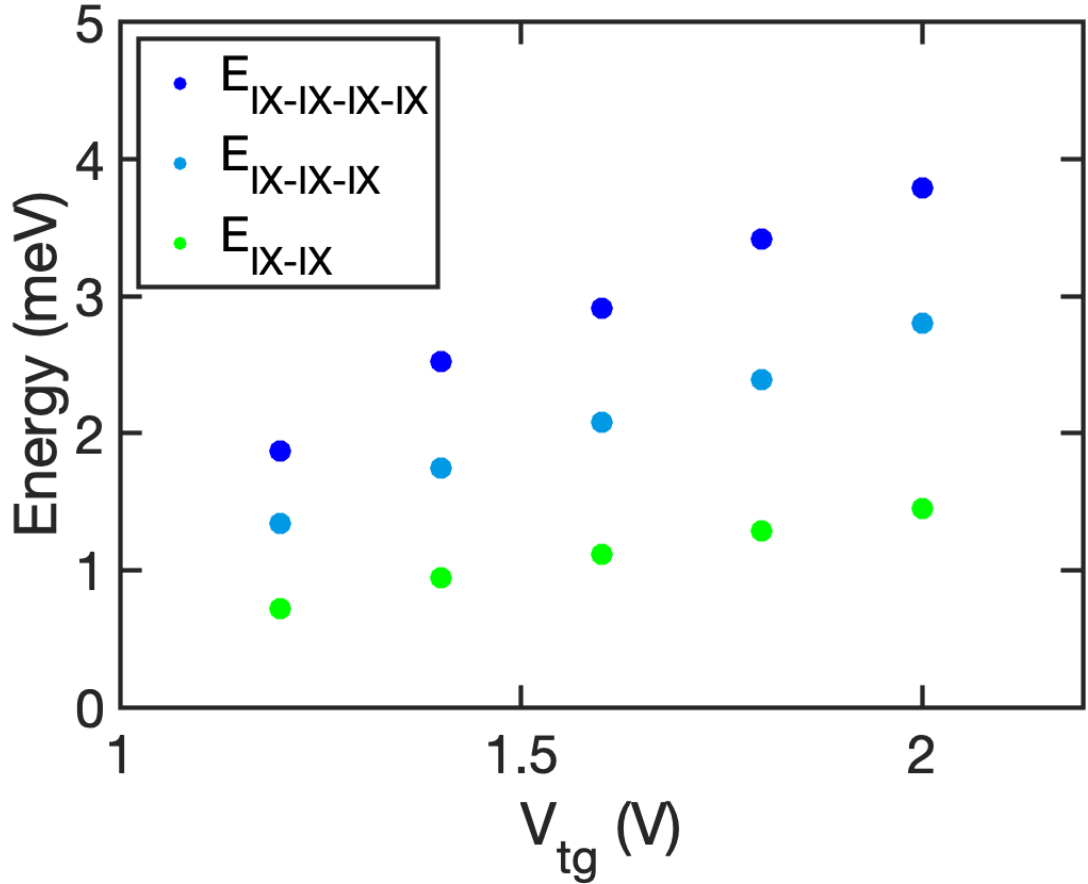

FIG. S5. Simulated multi-exciton energies for varying gate applied gate voltage, using the simulation described in the main text, showing the expected difference in energy shift between different occupation states, using the oblong shaped graphene etch profile in Fig. 4b. The slopes of these calculated values show that the N = 2, 3, and 4 exciton states should have dipole shifts that are 0.8, 1.7, and 2.6 meV/V larger than the single IX state, indicating that line *b* in Fig. S4 with a dipole shift 0.8 meV/V greater than line *a* is likely the bi-exciton state, and lines *c-e* with dipole shifts 5-10 meV larger than line *a* are likely multi-exciton states with N>4.

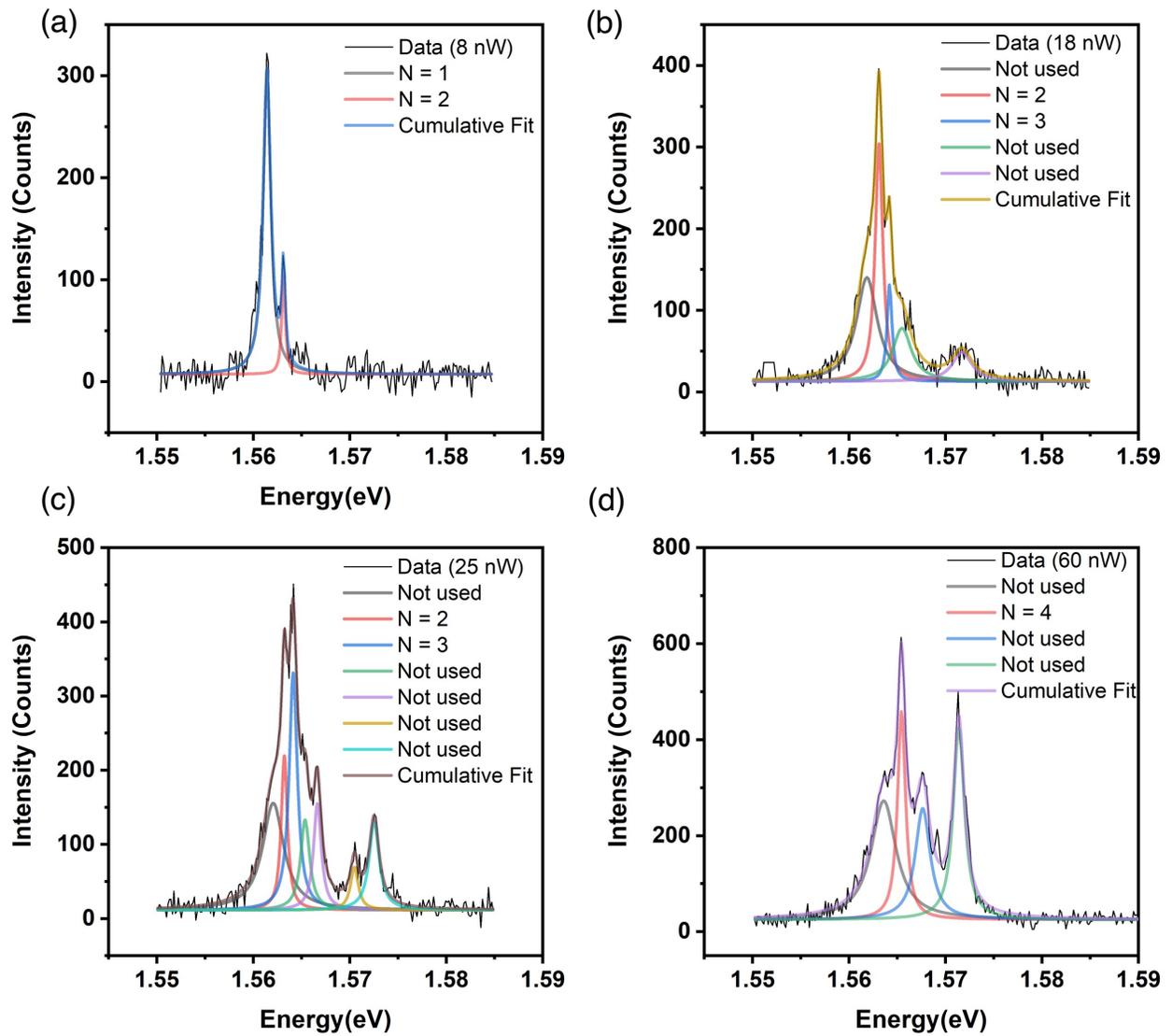

FIG. S6. Some representative fits to PL signals from Fig. 3, with center energies from fit used to determine energies for N = 1-4 states.

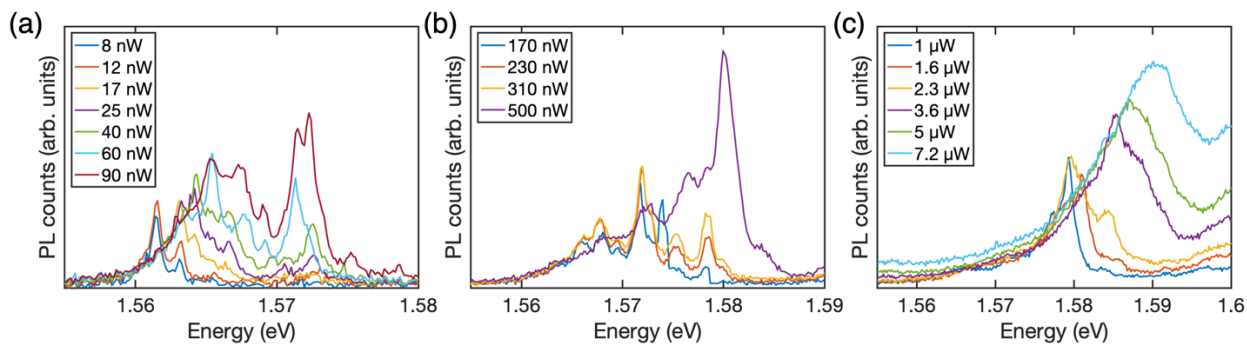

FIG. S7. Power dependent confocal PL of IXs in the electrostatic trap, showing the transition from narrow emission lines to a broad emission peak. We observe an IX energy shift of 30 meV between the lowest and highest excitation power PL scans. This is the largest power-induced IX blue-shift to our knowledge, showing that the funneling of IXs to the bottom of the electrostatic induced potential energy gradient plays a significant factor in IX dynamics. The PL data between $60 - 2300$ nW shows multiple sharp peaks, possibly indicating the presence of ordered IX-IX complexes at higher population. Such complexes will be the subject of future studies. The highest excitation powers measured, >3600 nW, we observe a broad emission peak, indicating that trapped IXs form a disordered state at high population.

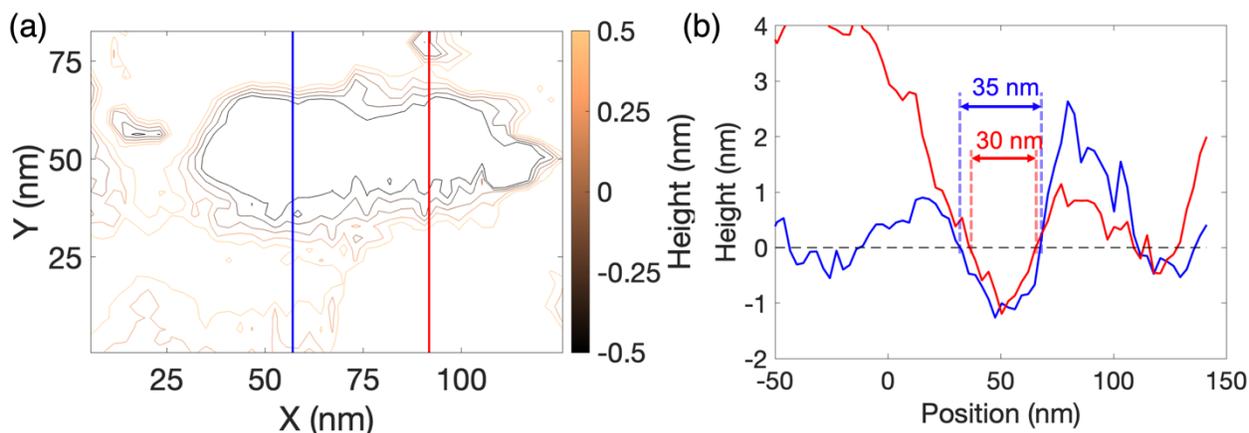

FIG. S8. AFM linecuts of the etched graphene. (a) Contour plot of the AFM topography with blue and red line cut locations shown. (b) Line cuts corresponding to (a), with a black dashed line at z = 0 nm height, showing a small difference in cross-sectional width of the hole on the left and right sides. The AFM scan is artificially flattened to first order in x and y, setting the flat, unetched areas on the structure to be of zero average height. We then measure zero-crossing points of the line cuts to be 35 nm (blue) and 30 nm (red). The exact shape of the hole is difficult to discern, given the AFM tip radius of 8 nm, and we additionally note that the AFM measurement underestimates the size of the hole due to the size of the tip radius. However, we note that small asymmetry in the shape of the graphene etch can produce significantly different multi-exciton energies, as shown in Table I.

**References:**

[1] P. Rivera et al., *Observation of Long-Lived Interlayer Excitons in Monolayer MoSe₂ –WSe₂ Heterostructures*, Nat Commun **6**, 1 (2015).

[2] J. R. Schaibley, P. Rivera, H. Yu, K. L. Seyler, J. Yan, D. G. Mandrus, T. Taniguchi, K. Watanabe, W. Yao, and X. Xu, *Directional Interlayer Spin-Valley Transfer in Two-Dimensional Heterostructures*, Nat Commun **7**, 1 (2016).

[3] J. R. Schaibley, H. Yu, G. Clark, P. Rivera, J. S. Ross, K. L. Seyler, W. Yao, and X. Xu, *Valleytronics in 2D Materials*, Nat Rev Mater **1**, 11 (2016).

[4] J. S. Ross et al., *Interlayer Exciton Optoelectronics in a 2D Heterostructure p–n Junction*, Nano Lett. **17**, 638 (2017).

[5] P. Rivera, H. Yu, K. L. Seyler, N. P. Wilson, W. Yao, and X. Xu, *Interlayer Valley Excitons in Heterobilayers of Transition Metal Dichalcogenides*, Nature Nanotech **13**, 1004 (2018).

[6] P. Nagler, F. Mooshammer, J. Kunstmann, M. V. Ballottin, A. Mitioglu, A. Chernikov, A. Chaves, F. Stein, N. Paradiso, S. Meier et al., *Interlayer Excitons in Transition-Metal Dichalcogenide Heterobilayers*, Physica Status Solidi (b) **256**, 1900308 (2019).

[7] F. Mahdikhanysarvejahany, D. N. Shanks, C. Muccianti, B. H. Badada, I. Idi, A. Alfrey, S. Raglow, M. R. Koehler, D. G. Mandrus, T. Taniguchi et al., *Temperature Dependent Moiré Trapping of Interlayer Excitons in MoSe₂-WSe₂ Heterostructures*, Npj 2D Mater Appl **5**, 1 (2021).

[8] W. Li, X. Lu, S. Dubey, L. Devenica, and A. Srivastava, *Dipolar Interactions between Localized Interlayer Excitons in van Der Waals Heterostructures*, Nat. Mater. **19**, 624 (2020).

[9] M. Kremser, M. Brotons-Gisbert, J. Knörzer, J. Gückelhorn, M. Meyer, M. Barbone, A. V. Stier, B. D. Gerardot, K. Müller, and J. J. Finley, *Discrete Interactions between a Few Interlayer Excitons Trapped at a MoSe₂–WSe₂ Heterointerface*, Npj 2D Mater Appl **4**, 1 (2020).

[10] D. N. Shanks, F. Mahdikhanysarvejahany, C. Muccianti, A. Alfrey, M. R. Koehler, D. G. Mandrus, T. Taniguchi, K. Watanabe, H. Yu, B. J. LeRoy et al., *Nanoscale Trapping of Interlayer Excitons in a 2D Semiconductor Heterostructure*, Nano Lett. **21**, 5641 (2021).

[11] I. Aharonovich, D. Englund, and M. Toth, *Solid-State Single-Photon Emitters*, Nature Photon **10**, 631 (2016).

[12] D. J. Morrow and X. Ma, *Trapping Interlayer Excitons in van Der Waals Heterostructures by Potential Arrays*, Phys. Rev. B **104**, 195302 (2021).

[13] *See Supplemental Material at [URL Will Be Inserted by Publisher] for Details on the Quantum Confinement Calculation, Keldysh Potential, Optical Image of the Device, Extended Power and Electric Field Dependent PL Data, PL Data from Alternate QD Sites, Calculated Multi-IX Energies as a Function of Electric Field, Representative Fits to PL Data, and Extended AFM Data of the Etched Graphene Hole.*

[14] D. Unuchek, A. Ciarrocchi, A. Avsar, Z. Sun, K. Watanabe, T. Taniguchi, and A. Kis, *Valley-Polarized Exciton Currents in a van Der Waals Heterostructure*, Nat. Nanotechnol. **14**, 1104 (2019).

[15] D. N. Shanks, F. Mahdikhanysarvejahany, T. G. Stanfill, M. R. Koehler, D. G. Mandrus, T. Taniguchi, K. Watanabe, B. J. LeRoy, and J. R. Schaibley, *Interlayer Exciton Diode and Transistor*, Nano Lett. **22**, 6599 (2022).

[16] F. Mahdikhanysarvejahany, D. N. Shanks, M. Klein, Q. Wang, M. R. Koehler, D. G. Mandrus, T. Taniguchi, K. Watanabe, O. L. A. Monti, B. J. LeRoy et al., *Localized Interlayer Excitons in MoSe$_2$–WSe$_2$ Heterostructures without a Moiré Potential*, Nat Commun **13**, 5354 (2022).

[17] N. Kumar, S. Najmaei, Q. Cui, F. Ceballos, P. M. Ajayan, J. Lou, and H. Zhao, *Second Harmonic Microscopy of Monolayer MoS$_2$*, Phys. Rev. B **87**, 161403(R) (2013).

[18] L. M. Malard, T. V. Alencar, A. P. M. Barboza, K. F. Mak, and A. M. de Paula, *Observation of Intense Second Harmonic Generation from MoS$_2$ Atomic Crystals*, Phys. Rev. B **87**, 201401(R) (2013).

[19] H. Zeng, G. B. Lui, J. Dai, Y. Yan, B. Zhu, R. He, L. Xie, S. Xu, X. Chen, W. Yao et al., *Optical Signature of Symmetry Variations and Spin-Valley Coupling in Atomically Thin Tungsten Dichalcogenides*, Sci Rep **3**, 1608 (2013).

[20] T. K. Stanev, P. Liu, H. Zeng, E. J. Lenferink, A. A. Murthy, N. Speiser, K. Watanabe, T. Taniguchi, V. P. Dravid, and N. P. Stern, *Direct Patterning of Optoelectronic Nanostructures Using Encapsulated Layered Transition Metal Dichalcogenides*, ACS Appl. Mater. Interfaces (2022).

[21] D. Yagodkin, K. Greben, A. E. Ascunce, S. Kovalchuk, M. Ghorbani-Asl, M. Jain, S. Kretschmen, N. Severin, J. Rabe, A. V. Krasheninnikov et al., *Extrinsic Localized Excitons in Patterned 2D Semiconductors*, Advanced Functional Materials **32**, 2203060 (2022).

[22] Z. Liu, Y. Gong, W. Zhou, L. Ma, J. Yu, J. C. Idrobo, J. Jung, A. H. MacDonald, R. Vajtai, J. Lou et al., *Ultrathin High-Temperature Oxidation-Resistant Coatings of Hexagonal Boron Nitride*, Nat Commun **4**, 2541 (2013).

[23] A. Laturia, M. L. Van de Put, and W. G. Vandenberghe, *Dielectric Properties of Hexagonal Boron Nitride and Transition Metal Dichalcogenides: From Monolayer to Bulk*, Npj 2D Mater Appl **2**, 1 (2018).

[24] L. A. Jauregui, A. Y. Joe, K. Pistunova, D. S. Wild, A. A. High, Y. Zhou, G. Scuri, K. De Greve, A. Sushko, C. H. Yu et al., *Electrical Control of Interlayer Exciton Dynamics in Atomically Thin Heterostructures*, Science **366**, 870 (2019).

[25] L. V. Butov, A. A. Shashkin, V. T. Dolgopolov, K. L. Campman, and A. C. Gossard, *Magneto-Optics of the Spatially Separated Electron and Hole Layers in GaAs/Al$_x$Ga$_{1-x}$As Coupled Quantum Wells*, Phys. Rev. B **60**, 8753 (1999).

[26] B. Laikhtman and R. Rapaport, *Exciton Correlations in Coupled Quantum Wells and Their Luminescence Blue Shift*, Phys. Rev. B **80**, 195313 (2009).

[27] Z. Sun, A. Ciarrocchi, F. Tagarelli, J. F. Gonzalez Marin, K. Watanabe, T. Taniguchi, and A. Kis, *Excitonic Transport Driven by Repulsive Dipolar Interaction in a van Der Waals Heterostructure*, Nat. Photon. **16**, 79 (2019).

[28] R. Rosati, S. Brem, R. Perea-Causín, R. Schmidt, I. Niehues, S. M. de Vasconcellos, R. Bratschitsch, and E. Malic, *Strain-Dependent Exciton Diffusion in Transition Metal Dichalcogenides*, 2D Mater. **8**, 015030 (2020).

[29] M. G. Harats, J. N. Kirchhof, M. Qiao, K. Greben, and K. I. Bolotin, *Dynamics and Efficient Conversion of Excitons to Trions in Non-Uniformly Strained Monolayer WS$_2$*, Nat. Photonics **14**, 5 (2020).

[30] T. P. Darlington, C. Carmesin, M. Florian, E. Yanev, O. Ajayi, J. Ardelean, D. Rhodes, A. Ghiotto, A. Krayev, K. Watanabe et al., *Imaging Strain-Localized Excitons in Nanoscale Bubbles of Monolayer WSe$_2$ at Room Temperature*, Nat. Nanotechnol. **15**, 10 (2020).

[31] A. Zrenner, *A Close Look on Single Quantum Dots*, J. Chem. Phys. **112**, 7790 (2000).

[32] J. J. Finley, A. D. Ashmore, A. Lemaître, D. J. Mowbray, M. S. Skolnick, I. E. Itskevich, P. A. Maksym, M. Hopkinson, and T. F. Krauss, *Charged and Neutral Exciton Complexes in Individual Self-Assembled In(Ga)As Quantum Dots*, Phys. Rev. B **63**, 073307 (2001).

[33] G. J. Schinner, J. Repp, E. Schubert, A. K. Rai, D. Reuter, A. D. Wieck, A. O. Govorov, A. W. Holleitner, and J. P. Kotthaus, *Confinement and Interaction of Single Indirect Excitons in a Voltage-Controlled Trap Formed Inside Double InGaAs Quantum Wells*, Phys. Rev. Lett. **110**, 127403 (2013).

[34] H. Yu, Y. Wang, Q. Tong, X. Xu, and W. Yao, *Anomalous Light Cones and Valley Optical Selection Rules of Interlayer Excitons in Twisted Heterobilayers*, Phys. Rev. Lett. **115**, 187002 (2015).

[35] H. C. Kamban and T. G. Pedersen, *Interlayer Excitons in van Der Waals Heterostructures: Binding Energy, Stark Shift, and Field-Induced Dissociation*, Sci Rep **10**, 5537 (2020).

[36] L. V. Keldysh, *Coulomb Interaction in Thin Semiconductor and Semimetal Films*, Soviet Journal of Experimental and Theoretical Physics Letters **29**, 658 (1979).

[37] A. Chernikov, T. C. Berkelbach, H. M. Hill, A. Rigosi, Y. Li, B. Aslan, D. R. Reichman, M. S. Hybertsen, and T. F. Heinz, *Exciton Binding Energy and Nonhydrogenic Rydberg Series in Monolayer WS$_2$*, Phys. Rev. Lett. **113**, 076802 (2014).

[38] T. C. Berkelbach, M. S. Hybertsen, and D. R. Reichman, *Theory of Neutral and Charged Excitons in Monolayer Transition Metal Dichalcogenides*, Phys. Rev. B **88**, 045318 (2013).

[39] W. Wang and X. Ma, *Strain-Induced Trapping of Indirect Excitons in MoSe$_2$/WSe$_2$ Heterostructures*, ACS Photonics **7**, 2460 (2020).

[40] G. Wei, D. A. Czaplewski, E. J. Lenferink, T. K. Stanev, I. W. Jung, and N. P. Stern, *Size-Tunable Lateral Confinement in Monolayer Semiconductors*, Sci Rep **7**, 1 (2017).

[41] M. Atatüre, D. Englund, N. Vamivakas, S.-Y. Lee, and J. Wrachtrup, *Material Platforms for Spin-Based Photonic Quantum Technologies*, Nat Rev Mater **3**, 38 (2018).

[42] D. E. Chang, A. S. Sørensen, E. A. Demler, and M. D. Lukin, *A Single-Photon Transistor Using Nanoscale Surface Plasmons*, Nature Phys **3**, 807 (2007).